\documentclass[twocolumn,showpacs,preprintnumbers,nofootinbib,prd,aps,
superscriptaddress,10pt]{revtex4-1}

\usepackage{amsmath,amssymb}
\usepackage[normalem]{ulem}
\usepackage{textcomp}
\usepackage{hyperref}
\usepackage{bm}
\usepackage{graphicx}
\usepackage{psfrag}
\usepackage[usenames,dvipsnames]{xcolor}
\usepackage[utf8]{inputenc}
\usepackage[english, capitalise]{cleveref}
\crefname{chapter}{Chap.}{Chap.}
\crefname{section}{Sec.}{Sec.}
\Crefname{chapter}{Chapter}{Chapters}
\Crefname{section}{Section}{Sections}
\Crefname{eqs}{Eqs.}{Eqs.}		
\Crefformat{eqs}{Eqs.~(#2#1#3)}	

\definecolor{darkgreen}{rgb}{0,0.5,0}
\hypersetup{
	bookmarks=true,         
	unicode=false,          
	pdftoolbar=true,        
	pdfmenubar=true,        
	pdffitwindow=false,     
	pdfstartview={FitH},    
	pdftitle={NSBH memory},
	pdfauthor={S.~Tiwari, M.~Ebersold, E.Z. Hamilton},
	pdfsubject={Gravitational Waves},
	pdfkeywords={gravitational waves, memory effect, general relativity},
	pdfnewwindow=true,      
	colorlinks=true,       
	linkcolor=red,          
	citecolor=cyan,        
	filecolor=magenta,      
	urlcolor=darkgreen,           
	linktocpage=true
}

\allowdisplaybreaks[4]

\newcommand{\Msun}{M_\odot}
\newcommand{\mem}{\textnormal{mem}}
\newcommand{\ri}{\mathrm{i}}

\newcommand{\BH}{\mathrm{BH}}
\newcommand{\NS}{\mathrm{NS}}

\begin{document}

\title{Leveraging gravitational-wave memory to distinguish neutron star-black hole binaries from black hole binaries}

\date{\today}

\author{Shubhanshu Tiwari}
\affiliation{Physik-Institut, Universit\"at Z\"urich, Winterthurerstrasse 190, 8057 Z\"urich, Switzerland}

\author{Michael Ebersold}
\affiliation{Physik-Institut, Universit\"at Z\"urich, Winterthurerstrasse 190, 8057 Z\"urich, Switzerland}

\author{Eleanor Z. Hamilton}
\affiliation{Physik-Institut, Universit\"at Z\"urich, Winterthurerstrasse 190, 8057 Z\"urich, Switzerland}

\begin{abstract} 
In the observation of gravitational waves (GWs) from a compact binary coalescence where the mass of one of the companions is less than $5~M_{\odot}$ the nature of the object is ambiguous until the measurements of tidal effects give evidence for the presence of a neutron star (NS) or a low mass black hole (BH). The relevance of tidal effects in a neutron star-black hole (NSBH) binary system depends crucially upon the mass and the spin of the companion BH. These effects become important predominantly when the binary system is of comparable mass and/or has large aligned spins. Depending upon the masses and spins the NS can even get tidally disrupted before the innermost stable circular orbit (ISCO) is reached. The gravitational-wave signatures of various tidal effects are encoded in the phasing of the signal and in the case of tidal disruption an abrupt cutoff of the signal amplitude occurs. In this work we show that tidal effects can also be captured by the nonlinear memory of the GW signal. Although small in amplitude, nonlinear memory is present at low frequency in contrast to the oscillatory GW signal. We introduce nonlinear memory in the NSBH and binary black hole (BBH) waveform models and show how the addition of memory aids in distinguishing NSBH systems from BBH systems for a larger part of the parameter space. We discuss the recently detected events of interest by LIGO-Virgo and provide the future prospects for the third generation detectors where nonlinear memory can play a crucial role in inferring the nature of the coalescence as BBH or NSBH from its GW signal alone.
\end{abstract}


\maketitle
\section{Introduction} \label{sec:introduction}
Neutron star-black hole binary systems are an extremely rich and versatile resource for gravitational wave astronomy. The recent detection of two NSBH binaries from their gravitational wave signal by the LIGO--Virgo detector network~\cite{LIGO-2014,VIRGO-2014} has provided evidence for their existence in our universe~\cite{LIGO-NSBH}. However, most of the confidence in the nature of these binaries comes from the measurement of the masses of the lighter companions, which are well within the mass range of known NS~\cite{antoniadis-2016,GW170817-2017,LIGO-BNS-2,alsing-2017}. Conclusive evidence for a companion being a NS through GW observation can only come from the measurement of tidal effects~\cite{binnington-2009,damour-2009,chatziioannou-2018,letiec-2020,charalambous-2021}. Unambiguous detection of a population of NSBH systems will have unique and complementary consequences for a wide range of topics such as astrophysics~\cite{broekgaarden-2021} and cosmology~\cite{vitale-2018}. It should also be noted that the detection of a BBH system where at least one of the companions has a mass similar to that of a NS ($\lesssim 3~\Msun $) will also have strong consequences and a clear hint toward new physics~\cite{LIGO-subsolar-2019}.   

The GWs from NSBH systems exhibit unique differences when compared to either binary neutron star (BNS) or BBH systems. \textit{First}, the tidal effects in the case of NSBH are not as pronounced compared to BNS systems as only one companion can evince tidal effects. This leads to an increase in difficulty measuring tidal effects. \textit{Second}, NSBH systems can go to very high mass ratios $m_\BH/m_\NS$ as the BH mass is not bounded, unlike BNS systems. With high mass ratio the tidal effects in NSBH systems become uninformative as the NS will plunge into the BH before having any substantial effect on the waveform~\cite{matas-2020}. For systems where the BH is only slowly spinning and has a substantially larger mass than the NS, the GW signal resembles a BBH system with same masses~\cite{foucart-2013}. This was the case for GW190814~\cite{abbott-2020-GW190814}. In this event a $23.2^{+1.1}_{-1.0}~\Msun$ BH coalesced with a $2.59^{+0.08}_{-0.09}~\Msun$ compact object, whose nature is unknown. It could be either the heaviest NS or the lightest BH observed in a compact binary system. \textit{Lastly}, for NSBH systems where the companions are of similar masses or the BH has high positive spin, finite size effects cause a dephasing of the waveform relative to a BBH with same masses and spins while tidal forces on the NS affect the amplitude of the waveform~\cite{flanagan-2007,vines-2010}. Depending on its equation of state, the NS can be entirely torn apart in a tidal disruption event~\cite{kyutoku-2010,kyutoku-2011,foucart-2012}. Such an event can be accompanied by the formation of a torus remnant around the BH~\cite{foucart-2012-2,lovelace-2013} and a kilonova~\cite{metzger-2016,tanaka-2016}, if the tidal disruption happens before the innermost stable circular orbit (ISCO). Then also the GW signal is highly affected, the amplitude of the merger is heavily suppressed and the ringdown part is missing completely~\cite{pannarale-2015,kawaguchi-2017}. As the tidal disruption event can only happen for NSBH systems it would be a smoking gun evidence for a NSBH binary. However, it should be noted that the disruption frequency will be in the frequency range of a few kHz where the detectors' sensitivity is not optimal~\cite{matas-2020,aLIGO-o3Performance-2020}. The scenario of tidal disruption is not only the best case for the unambiguous identification of NSBH system but also can lead to potential electromagnetic counterparts~\cite{metzger-2016,foucart-2016,barbieri-2019,foucart-2020,fragione-2021}. In this work we motivate the use of gravitational-wave memory as a complementary resource for identifying the tidal disruption scenario which will help in distinguishing NSBH from BBH mergers. 

The gravitational memory effect predicts a persistent physical change to spacetime induced by the passage of transient gravitational radiation. Therefore, the proper distance between the locations of freely falling observers differ before and after a GW passes through. There are two kind of memory effects: the first is a linear memory effect associated typically with components or matter being unbound, e.g. binaries on hyperbolic orbits~\cite{zeldovich-1974,braginsky-1987,turner-1978}. The second kind, the nonlinear memory effect is directly related to the nonlinearity of general relativity~\cite{christodoulou-1991,thorne-1992,blanchet-1992}. Here we concentrate on the type of nonlinear memory known as Christodoulou memory or displacement memory, which is the most prominent kind of memory present in GWs from bound compact binary systems, hence we will refer to it from here on as just memory. This additional nonoscillatory component to a GW signal can be understood as being sourced by the traveling GWs themselves. The GW memory effect also has close connections to the symmetry group of asymptotically flat spacetimes, the Bondi-Metzner-Sachs (BMS) group~\cite{bondi-1962,sachs-1962,sachs-1962-2}, and its corresponding conserved quantities (see, e.g., Ref.~\cite{strominger-2017} for more details). Several studies have been performed on the prospect of detecting memory with LIGO, LISA or other future GW detectors~\cite{favata-2010,lasky-2016,johnson-2018,islo-2019,boersma-2020,huebner-2020,huebner-2021}.

The memory component of GWs from compact binary coalescences (CBCs) slowly accumulates during the inspiral and significantly jumps during the merger phase of the evolution when spacetime curvature is highest. This adds a late-time low frequency component to the waveforms of CBC systems. This occurs at the point in the signal where the frequencies go beyond the sensitive range of current generation ground-based detectors. This low frequency component has been used to look for sources that emit GWs at very high frequencies~\cite{mcneill-2017}, for example very light BBH mergers~\cite{ebersold-2020}, thus providing a creative way to widen the parameter space of GW searches.
In this context, the utility of memory for detection and interpretation of sources can become significant when the purely oscillatory part of the signal extends to such high frequencies that the 
detectors become less sensitive. Tidal disruption of neutron stars in NSBH systems is likely to happen at a few kHz and therefore beyond the sensitive spot of the detectors~\cite{pannarale-2015,aLIGO-o3Performance-2020}. We show that in this case the memory can provide crucial information about the nature of the system.

The rest of the paper is organized as follows: In Sec.~\ref{sec:memory} we describe the waveform model we use, which includes memory for the case of NSBH and BBH coalescences. We then discuss the parameter space where memory can be utilized to identify the tidal disruption event in Sec.~\ref{sec:tidal}. In Sec.~\ref{sec:prospects} we quantify the amount to which memory can contribute toward the distinguishability of NSBH and BBH mergers. We discuss the current events of interest in Sec.~\ref{sec:events} and finally give our conclusions in Sec.~\ref{sec:conclusion}.

\section{Memory of neutron star-black hole systems} \label{sec:memory}
Current state-of-the-art waveforms used in GW data analysis do not contain memory.
However, the contribution from the memory effect to the gravitational waveform can be computed 
from any waveform in the time-domain~\cite{favata-2009,talbot-2018,ebersold-2019,ebersold-2020}. It is convenient to decompose the GW polarizations into modes $h_{\ell m}$ via
\begin{align}
h_+(t) - \ri h_\times(t) = \sum_{\ell = 2}^{\infty} \sum_{m = -\ell}^{\ell} h_{\ell m}(t)\, {}_{-2}Y_{\ell m}(\iota, \Phi)\,,
\end{align}
where the basis ${}_{-2}Y_{\ell m}(\iota, \Phi)$ is formed by the spin-weighted spherical harmonics 
with spin weight -2. The angles $\iota$ and $\Phi$ denote inclination and a reference phase of the source (typically the phase at coalescence for compact binaries). 
For nonprecessing binaries the memory contribution can be assumed to be linearly polarized and only appears in the plus polarization. Additionally, considering only the dominant oscillatory waveform mode $h_{22}$, the general formula for the memory~\cite{ebersold-2020} simplifies to:
\begin{align} \label{eq:mem}
	h_+^\mem(T_R) = \frac{R}{7c} \sqrt{\frac{5}{6\pi}} \int^{T_R}_{-\infty} dt \, |\dot h_{22}(t)|^2\, {}_{-2}Y_{2 0}(\iota, \Phi) \,,
\end{align}
where the integral from past infinity is taken from the start of the oscillatory waveform to retarded time $T_R$. The overdot represents the time derivative and $R$ is the distance to the source.
Neglecting contributions from higher order oscillatory modes 
typically leads to underestimating the memory amplitude by up to $10\%$. Higher order memory modes for the case of nonprecessing binaries are negligible~\cite{talbot-2018}.  

\begin{figure}[h]
  \centering
  \scalebox{0.48}{\includegraphics{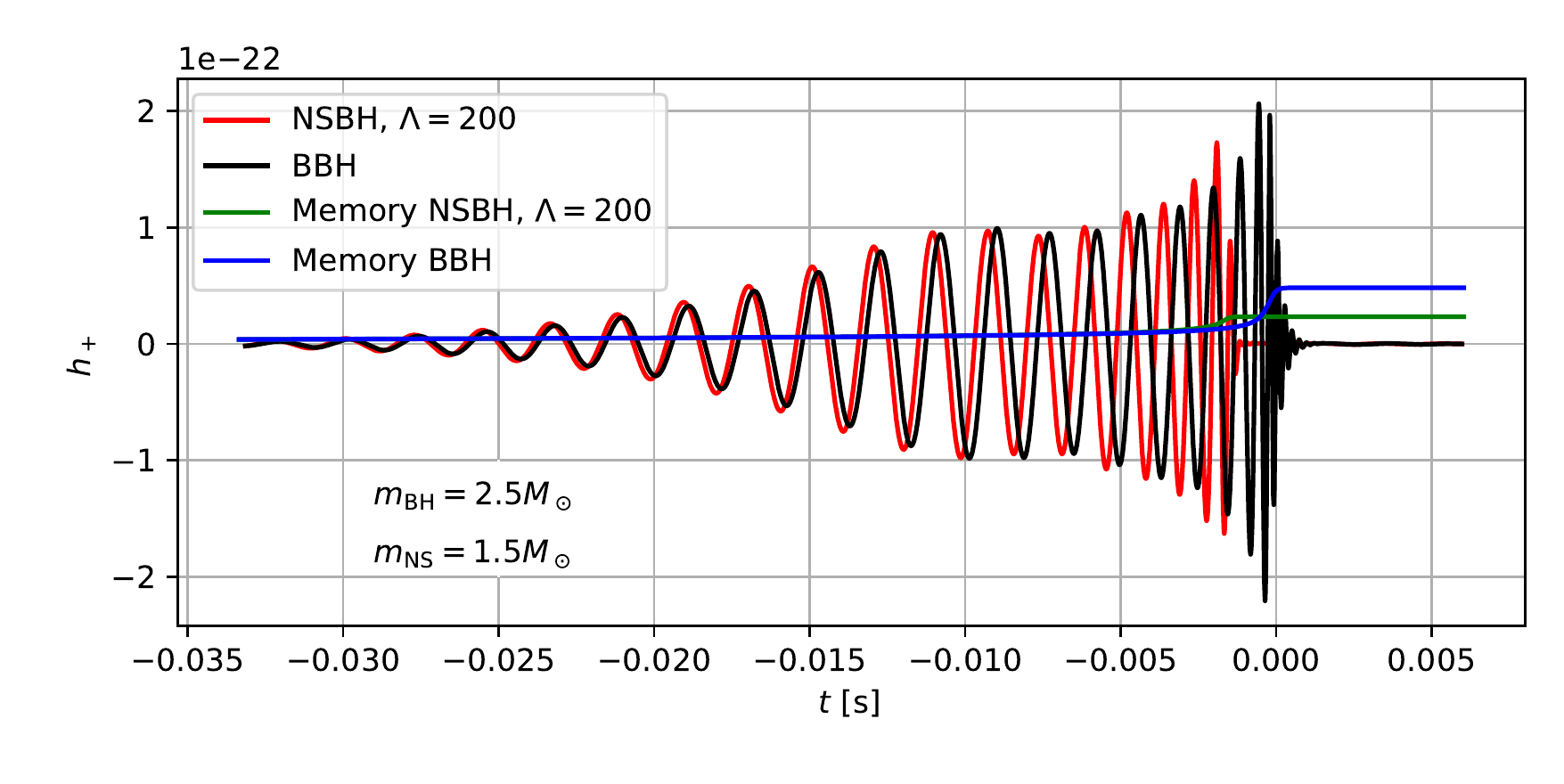}}
  \caption{The oscillatory GW amplitude tapered at 500~Hz of an NSBH and a corresponding BBH coalescence at a distance of 100~Mpc and at an inclination angle of $\iota = 90^\circ$. We also show the memory contribution of the two systems.}
  \label{fig:oscandmem}
\end{figure}

The waveform model we use for the oscillatory waveform from aligned-spin NSBH systems is called 
\texttt{SEOBNRv4\_ROM\_NRTidalv2\_NSBH}~\cite{matas-2020}. It is a frequency-domain model based on the BBH model \texttt{SEOBNRv4}~\cite{bohe-2016} relying on the effective-one-body (EOB) framework~\cite{buonanno-1998,dietrich-2017}. Tidal effects on the phase are incorporated from Ref.~\cite{dietrich-2019} and the amplitude is corrected to account for tidal disruption inspired by Ref.~\cite{pannarale-2015} as well as numerical relativity (NR) simulations of NSBH coalescences~\cite{matas-2020}.
The model takes five parameters, the respective masses and dimensionless aligned spins of the BH and the NS, $M_\BH$, $M_\NS$, $\chi_\BH$ and $\chi_\NS$, and the dimensionless tidal deformability parameter of the NS defined as 
\begin{align}
    \Lambda_\NS = \frac{2}{3} k_2 \frac{R_\NS^5}{M_\NS^5} \,,
\end{align}
where $k_2$ is the Love number and $R_\NS$ is its radius~\cite{chatziioannou-2020}. Henceforth, we assume both spins to be zero and only comment on the impact of nonzero spins later on. If $\Lambda_\NS = 0$, the model reduces to a BBH waveform model. Since the memory contribution is computed in the time-domain, we call the NSBH waveform model in the time-domain via the LIGO Algorithmic Library (LAL) Suite~\cite{LAL-2021}. \cref{fig:oscandmem} shows the oscillatory waveform above 500~Hz for a binary system with component masses 2.5 and 1.5~$\Msun$. The primary objects is a BH and for the secondary object we consider two cases: first we consider an NS with $\Lambda_\NS = 200$ and second a BH. One can clearly distinguish the typical inspiral-merger-ringdown signal in the BBH case, whereas the merger-ringdown part is heavily suppressed in the NSBH case due to tidal disruption. Knowing the oscillatory waveform we can compute the dominant memory contribution according to~\cref{eq:mem}, which is also shown for both cases in~\cref{fig:oscandmem}. The jump in the memory is considerably higher in the BBH case.

\begin{figure}[t]
  \centering
  \scalebox{0.48}{\includegraphics{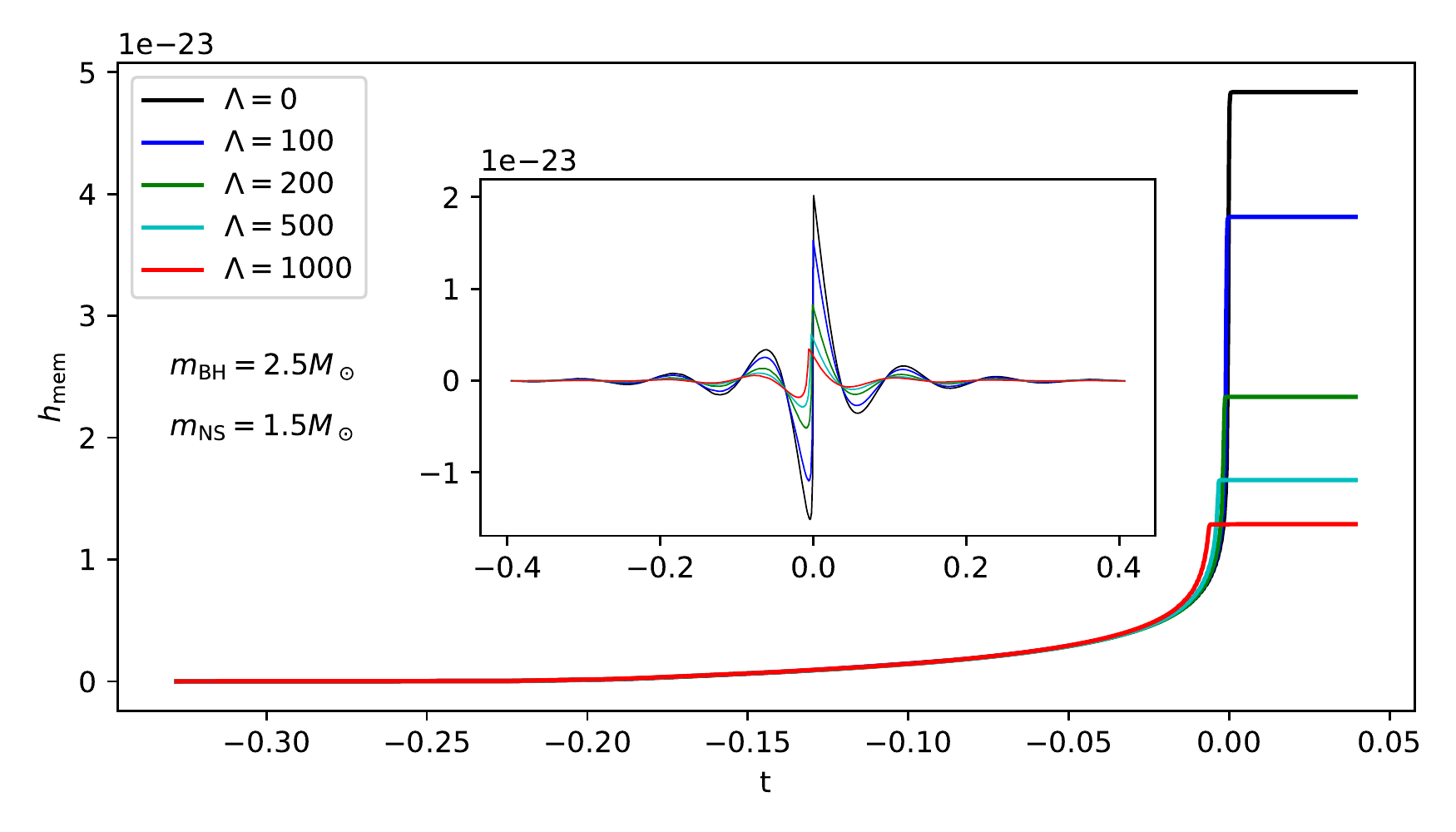}}
  \caption{The GW amplitude of the memory contribution from a NSBH system at 100~Mpc. Shown is the last part of the inspiral, from a frequency of 200~Hz until the merger. $\Lambda = 0$ corresponds to a BBH system. The inset shows the same GW signals but with a low-frequency cutoff at 10~Hz.}
  \label{fig:memamplitude}
\end{figure}

Figure~\ref{fig:memamplitude} shows the amplitude of the memory contribution for different values of $\Lambda_\NS$. The memory accumulated in the inspiral and merger is largest for a BBH and gets substantially smaller with increasing values of $\Lambda_\NS$. Here the memory is computed from the point where the inspiral frequency passes 200~Hz. The contribution from earlier times is neglected since its effect on the memory amplitude is insignificant.
A GW detector is only sensitive above a certain frequency and will therefore not see the eventual offset in the strain, but rather the rise in memory since that also contains nonzero frequencies. By applying a low-frequency cutoff with a high-pass filter we see how the memory is observed in the detector. We choose the cutoff frequency to be 10~Hz since that is the lower end of advanced LIGO's design sensitivity~\cite{aLigo-design-sens} (zero detuned high power configuration of the detectors). The resulting signal is a burst at the time of the merger, as shown in the inset of~\cref{fig:memamplitude} for a NSBH system with different values of the NS tidal deformability. 

As a cross check for our memory calculation according to~\cref{eq:mem} we apply the method developed in Ref.~\cite{mitman-2020-2} directly to NR waveforms of NSBH mergers by the SXS collaboration~\cite{foucart-2018}. Although it was stated earlier that extracting memory from NR simulations is possible~\cite{pollney-2010}, it is dependent on the extraction method and usually just neglected. Only recently it was computed directly from numerical relativity~\cite{mitman-2020} and shown how it can be added to waveforms in the SXS catalog of NR waveforms~\cite{boyle-2019,mitman-2020-2}.
Essentially by exploiting the BMS balance laws, the GW strain can be corrected to include the displacement memory. For the available nonspinning simulations \texttt{SXS:BHNS:0001}, \texttt{SXS:BHNS:0002}, \texttt{SXS:BHNS:0004} and \texttt{SXS:BHNS:0006}~\cite{SXS:BHNS:0001,SXS:BHNS:0002,SXS:BHNS:0004,SXS:BHNS:0006}, which were also used to tune the model \texttt{SEOBNRv4\_ROM\_NRTidalv2\_NSBH}~\cite{matas-2020}, we find no major discrepancies in the memory amplitude with both approaches. Although, a systematic comparison would go beyond the scope of this paper.

The amplitude of the memory as compared to the oscillatory part of the gravitational 
wave signal is much smaller, typically about an order of magnitude. For fixed total mass, the memory amplitude is maximal in equal mass systems and decreases with more unequal masses.
Moreover, it has a distinct dependence on the inclination angle of~\cite{talbot-2018}
\begin{align}
    h_+^\mem \sim \sin^2 \iota \left( 17 + \cos^2 \iota \right)\,,
\end{align}
thus it is maximal for edge-on systems and vanishes for face-on systems.
However, it has a completely different frequency content than the oscillatory inspiral-merger-ringdown signal. It is basically just a burst saturating toward the lower frequencies at the time of the merger. 

\section{Nonlinear memory as an identifier of tidal disruption event} \label{sec:tidal}
Identifying the nature of the compact objects in a CBC from their GW signal alone where one of the companions is low mass $< 5~M_{\odot}$ is a challenging task. For example, we have seen the detection of the events GW190814~\cite{LIGO-190814}, GW190425~\cite{LIGO-BNS-2}, GW200105 and GW200115~\cite{LIGO-NSBH}, where at least one of the components was low mass and without an electromagnetic (EM) counterpart, the nature of the lighter companion was not well established. The \textit{prima facie} reason for this challenge can be largely attributed to the fact that we rely on the measurement of tidal deformation parameters different from null. Measuring tidal deformation parameters is difficult as these effects are small and often become more prominent at higher frequencies where the detectors' sensitivities are not optimal~\cite{matas-2020,hinderer-2009,lackey-2014}. Inclusion of memory provides a complementary way to infer the nature of the less massive compact object.

In this section we start by summarizing the different morphologies of NSBH binaries following Refs.~\cite{pannarale-2015,matas-2020} and discuss how the memory relates with each of these cases. Then we briefly discuss the impact of linear memory generation and the influence of an aligned BH spin on the memory and a potential tidal disruption event. Later we present the case for when memory is most useful for identifying the nature of compact objects in a low-mass CBC signal when EM counterparts are not detected.

\subsection{Memory signal for various cases of tidal disruptions }

\begin{figure}[t]
  \centering
  \scalebox{0.53}{\includegraphics{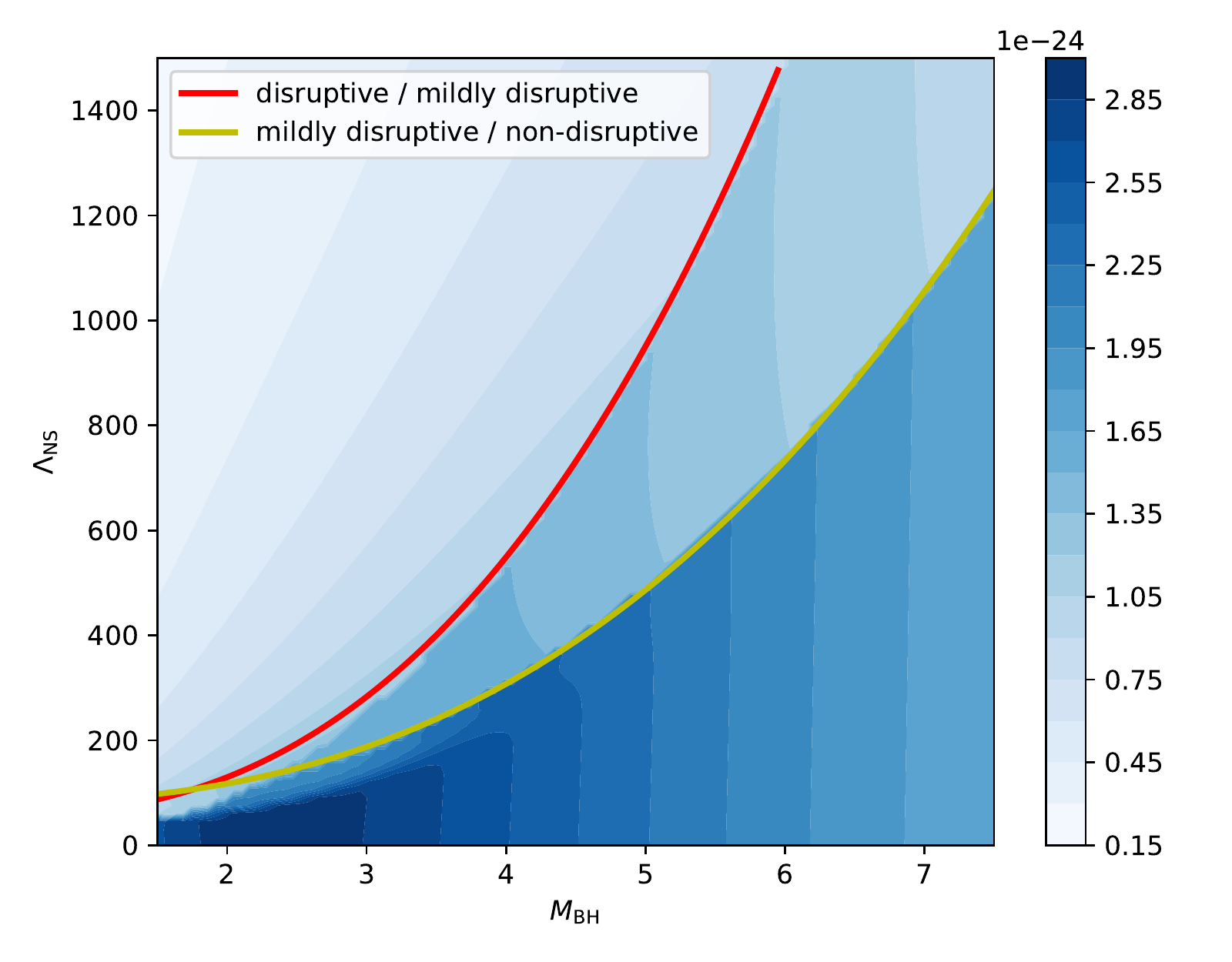}}
  \caption{The $h_{rss}$ of the memory amplitude for edge-on NSBH binaries is plotted as a function of the NS tidal deformability and the mass of the BH. The red (yellow) curve corresponds to the light- (dark-) gray region of the top right plot in Fig.~2 of Ref.~\cite{matas-2020}. The region below the yellow curve contains nondisruptive mergers, in between the curves there are mildly disruptive mergers without a torus remnant and above the red curve we find disruptive mergers with a torus remnant. The mass of the NS is fixed to $1.5~\Msun$ and the luminosity distance is 100~Mpc.}
  \label{fig:memhrss}
\end{figure}

In an NSBH inspiral due to gravitational wave emission, the BH exerts tidal forces on the NS. If the NS approaches sufficiently close to the BH that the tidal forces overcome the self gravity of the NS, it can lose mass in a process called mass shedding. This can lead to the tidal disruption of the NS, where it is completely torn apart due to the strong gravitational field of the BH. Depending on whether the mass shedding begins before the innermost-stable circular orbit (ISCO), after which the NS plunges into the BH and finally merges, the characteristic imprint on the emitted GW signal will be different. The fate of the NS depends mainly on its equation of state (and therefore the tidal deformability), the mass ratio and the spin of the BH. NSBH coalescences can be roughly classified in three categories~\cite{pannarale-2015,matas-2020}:
\begin{itemize}
\item[(i)] \textbf{Nondisruptive mergers:} The NS crosses the ISCO before mass shedding occurs and it plunges as a whole into the BH. The waveform looks similar to a BBH but with a slightly suppressed amplitude in the merger and ringdown.

\item[(ii)] \textbf{Mildly disruptive mergers:} Although the NS undergoes mass shedding prior to crossing the ISCO, it is not completely torn apart. In the GW signal the merger and ringdown phase is suppressed but still present.

\item[(iii)] \textbf{Disruptive mergers:} In this case the NS gets tidally disrupted before reaching the ISCO and a remnant torus of matter forms. The GW signal lacks a clear merger and ringdown part, instead the amplitude tapers off. 
\end{itemize}
The amplitude of the memory signal is largely correlated with the oscillatory waveform as can be seen in ~\cref{fig:memhrss}. The nature of the coalescence determines the morphology of the oscillatory waveform and therefore also affects the memory amplitude. For the first case of nondisruptive mergers, where the oscillatory waveform is BBH-like, the memory is more pronounced. This is because most of the memory is generated when the gravitational field is the strongest and therefore, like the oscillatory signal, the memory signal looks almost as in the case of a BBH with the same masses. For the other two cases where tidal disruption occurs, the lack of the merger-ringdown part of the signal means there is less overall contribution to the generation of memory, causing it to be much weaker. It should be noted that although the amplitude of the memory signal corresponds to the peak amplitude of the oscillatory signal the frequency spectra of the memory signal is completely different. When the CBC system consists of two low mass companions the oscillatory waveform can peak at frequencies beyond the sensitivities of the detectors but the memory signal will always appear at the low frequency cutoff of the detector. The correlation of the memory with the nature of the NSBH is shown in~\cref{fig:memhrss}, where we plot the root-sum-squared amplitude $h_{rss}$ of the memory waveform for a fixed NS mass of $1.5~M_\odot$ and with varying tidal deformability parameter $\Lambda_\NS$ and mass of the BH $M_\BH$. The different regimes can be clearly distinguished, the borders between disruptive/mildly disruptive and mildly disruptive/nondisruptive are indicated with red and yellow lines respectively. 

In the case of disruptive mergers, a substantial amount of matter can be ejected. A fraction of this matter can become unbound at relativistic velocities, which in turn can lead to the production of linear memory with an amplitude proportional to the amount of dynamically ejected mass. Assuming the most extreme configuration favorable to the production of linear memory, which is that all ejecta is moving away from the remnant as a point mass of $0.1~\Msun$ and velocity of $0.3~c$~\cite{foucart-2012,kyutoku-2015}, the amplitude of the linear memory computed following Ref.~\cite{favata-2009} can be of a similar order of magnitude as the nonlinear memory. Even in this case the build-up time of the linear memory is much slower than, and not comparable to, the jump of the nonlinear memory close to merger, thus it will be mostly discarded by the low-frequency cutoff. In a realistic scenario, the contribution to linear memory will only occur from the asymmetry in the distribution of ejecta, which would render the mass contributing to the production of linear memory much smaller. Another mechanism known to cause linear memory is the recoil or kick that the remnant receives due to anisotropic radiation of gravitational radiation, though its amplitude is generally at least two orders of magnitude smaller than the nonlinear memory as shown in Ref.~\cite{favata-2008}.

Considering a spin component of the BH that is aligned with the orbital angular momentum of the binary, tidal disruption occurs for a wider region in the parameter space as can be seen in Fig.~2 of Ref.~\cite{matas-2020}. In contrast for anti-aligned-spins tidal disruption is less favored. The reason is that aligned-spin systems tend to stall out the coalescence whereas antialigned systems are less stable and hence they plunge faster into each other~\cite{campanelli-2006}. We recall that memory has a similar dependence on the spin, due to the longer coalescence time for aligned-spin systems, more memory amplitude is accumulated~\cite{talbot-2018,ebersold-2020}.

\subsection{Complementing the oscillatory signal with memory} \label{comp}

The tidal deformability does not considerably affect the oscillatory waveform below $\sim 500$~Hz~\cite{Pannarale:2011pk,hotokezaka-2016} in a typical NSBH inspiral. Beyond this rough threshold a higher value of the tidal deformability leads to a faster inspiral and depending on the mass ratio of the system the NS could get tidally disrupted. Typical frequencies where tidal disruption can occur are in the kHz-range. This frequency is due to the fact that canonically the maximum mass of the NS will be less than $3~\Msun$~\cite{rhoades-1974,GW170817-2018,cromartie-2019,fonseca-2021,nathanail-2021} while the BH mass is likely to be above this limit. Since tidal disruption is favored for more equal mass systems we assume $M_\BH \sim \mathrm{a\;few}\; \Msun$. Since tidal disruption is a late stage inspiral phenomenon, the low total mass of the system means the frequency at which it occurs will be high. However, at higher frequencies the sensitivity of current generation detectors starts to fall off. We note that there are proposals for GW detectors which will be sensitive at high frequencies like~\cite{NEMO-2020} for which the oscillatory signal might be fully in band during the late inspiral and merger. As mentioned previously, the memory signal will occur at low frequencies and thus can provide additional information. Depending on whether the NS was tidally disrupted, swallowed by the BH as a whole or the system is actually a BBH, the burst of memory in the low frequencies is significantly different.

The dependence of the memory signal on the orientation or inclination angle of the binary is out of phase with the dependence of the oscillatory signal; the memory is maximal for an edge-on system while the oscillatory signal peaks when the system is face-on. This feature of the memory signal becomes very useful when we have events with low mass companions and with similar masses (mass ratio closer to 1) without any EM counterpart detection. If the event is accompanied by EM radiation the nature of the coalescence can be deduced independently from the GW signal. An NSBH merger is expected to be accompanied by a short gamma ray burst (GRB) if the NS gets tidally disrupted outside the BH's horizon~\cite{berger-2013}. This is due to the fact that more mass is ejected during the tidal disruption of a NS as compared to the case when the NS is not tidally disrupted~\cite{foucart-2015}. This can lead to a jet formation which is expected to be oriented perpendicular to the plane of the binary’s orbit. Thus a GRB would most likely be observed if the NSBH is face-on~\cite{gompertz-2020}. In that case the help of memory is minimal and the nature of the binary could be established by the presence/absence of a GRB. In contrast if the system is edge-on only the GW signal can be observed and memory will be maximum.
So the maximum contribution from memory happens when the event is expected not to have any other means to infer the nature of the coalescence apart from its GW signal.  
It should be noted that in addition to the short GRBs, NSBH mergers can also power kilonovae, a bright short lived quasi thermal emission mostly in infrared and optical wavelengths~\cite{metzger-2010,coulter-2017} and delayed afterglow mostly in x-ray and radio energies~\cite{hallinan-2017,troja-2017}. These EM counterparts can be visible for off-axis mergers and can probe the nature of the compact objects even if the orbital plane is face-off and the GRB jet is not facing the observer. However, the detection of kilonovae and afterglow is challenging without a precise sky direction information of the source which is at best of the order of a few square degrees from GW detectors~\cite{KAGRA:2013rdx}. 
In the case when the GRB is off-axis, memory plays a crucial role in identifying if the system can be a potential candidate to follow up potential kilonovae and afterglow by flagging the event as tidally disrupted NSBH or not.

\section{Prospects for identifying tidal disruption events using nonlinear memory} \label{sec:prospects}
In this section we present the parameter space where memory aids in distinguishing a NSBH binary merger undergoing tidal disruption from a BBH merger and quantify the degree to which memory can be useful. To do so we work with a metric which is widely used to measure how much two waveforms differ known as the match. To compute the match between different waveform templates we define the overlap or inner product between two templates $h$ and $g$ as~\cite{sathyaprakash-1991,finn-1992}
\begin{align}
    \langle h | g \rangle := 4 \mathrm{Re} \int df \frac{\bar h(f) g(f)}{S_n(f)}\,.
\end{align}
$S_n(f)$ denotes the power spectral density (PSD) of the detector noise.  
The match $\mathcal M$ is then computed by maximizing the overlap over coalescence time $t_c$ and phase of the templates $\phi_c$~\cite{damour-1998},
\begin{align}
    \mathcal M (h,g) = \max_{t_c, \phi_c} \left[ \frac{\langle h | g \rangle }{\sqrt{\langle h | h \rangle \langle g | g \rangle }} \right]\,.
\end{align}
We compute the match between different NSBH and BBH waveform templates and investigate how the addition of memory affects the match. The waveforms used here are the ones described in Sec.~\ref{sec:memory}. 

\begin{figure*}[t!]
  \centering
  \scalebox{0.8}{\includegraphics{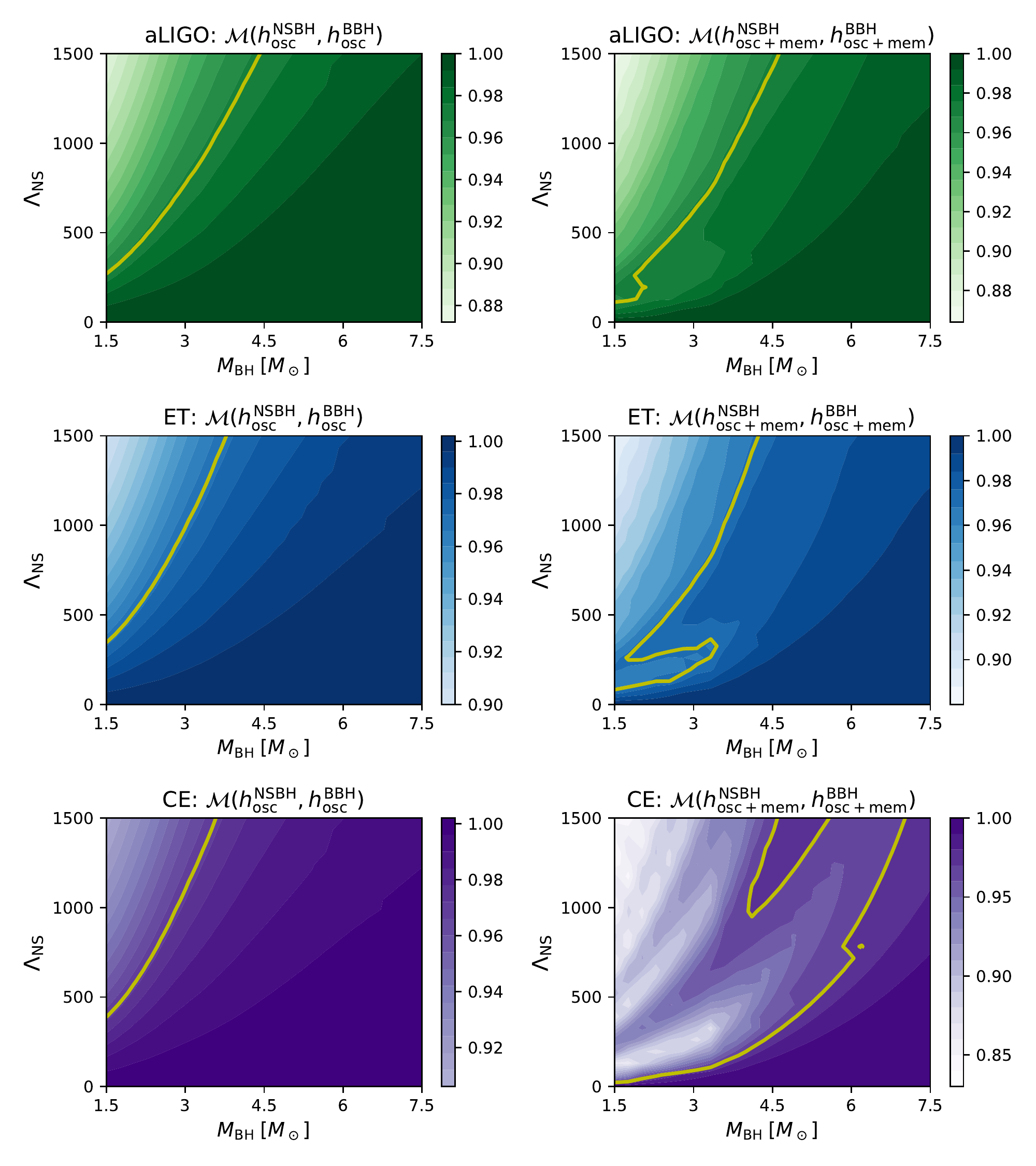}}
  \caption{The plots on the left hand side show the match of the GW signal between the oscillatory waveform in the latest stages of the inspiral  above 500~Hz and the merger/plunge from a NSBH and a corresponding BBH for different mass ratios and tidal deformabilities, these results are consistent with results obtained in~\cite{Pannarale:2011pk}. The plots on the right hand side show the same but including the memory contribution. The mass of the secondary object is fixed to $1.5~\Msun$. We display the match for different PSDs of current and future detectors, starting with advanced LIGO~\cite{aLigo-design-sens} on top, the Einstein Telescope~\cite{EinsteinTelescope-2019} in the middle and Cosmic Explorer~\cite{CosmicExplorer-2017} at the bottom. The oscillatory waveform is only taken into account above a frequency of 500 Hz as only there the waveforms start to differ according to Ref.~\cite{hotokezaka-2016,Pannarale:2011pk}. A yellow line marks a match of 0.97, where waveform templates can be distinguished to 90\% confidence if the event has an SNR of~10.}
  \label{fig:matchplot}
\end{figure*}

The resulting matches are presented in~\cref{fig:matchplot} for a range of simulated detector noise PSDs, namely advanced LIGO at design sensitivity~\cite{aLigo-design-sens}, Einstein Telescope (ET)~\cite{EinsteinTelescope-2019} and Cosmic Explorer (CE)~\cite{CosmicExplorer-2017}. The mass of the NS is always fixed to $1.5~\Msun$ and we vary the mass of the BH $M_\BH$ and the tidal deformability $\Lambda_\NS$ of the NS. For the BBH template the tidal deformability is of course always zero, so in the limit of $\Lambda_\NS \rightarrow 0$ we should always find a match of~1. The oscillatory templates start at 500~Hz where the effect of tidal deformation starts to have a major impact and a high-pass filter is applied to the memory waveforms in order to cut frequencies below 10~Hz. All waveforms are generated at an inclination angle of $90^\circ$ (edge-on orientation) where the effect of memory is maximized and also it is the case where the addition of memory will be the most beneficial, as discussed in Sec.~\ref{comp}. The plots on the left hand side show the match between only the oscillatory waveform templates. We can clearly see that the match falls for larger values $\Lambda_\NS$ and for more equal mass systems. This is expected as in this parameter space the systems can get tidally disrupted and therefore lack the final merger-ringdown part of the waveform compared to a corresponding BBH waveform. Otherwise the more unequal the masses of the two objects get and the smaller $\Lambda_\NS$, the waveform becomes more BBH-like and therefore the match remains high and it is harder to distinguish the systems.

The plots on the right of~\cref{fig:matchplot} show the matches but with the addition of memory to both the waveform templates. We find that lines of constant match are generally shifted toward the bottom right compared to the plots on the left. This is as anticipated, since systems that are tidally disrupted have much less memory amplitude compared to a corresponding BBH system. Moreover, we see that adding memory especially improves the distinguishability for small values of $\Lambda_\NS$ in almost equal mass binaries. Measured values of $\Lambda_\NS$ from the observation of GW170817~\cite{GW170817-2017}, which had a mass ratio of less than 1.5, show support for $\Lambda_\NS$ around 300 and $\Lambda_\NS \lesssim 600$ for even the broadest prior~\cite{GW170817-2018}, this is the region of $\Lambda_\NS$ where memory contributes the most for increasing the distinguishability. This is due to the fact that if one companion is a NS, these systems are likely to be disrupted, producing a dip in memory, whereas if both are BHs the memory amplitude benefits from the equal masses and peaks. 
As displayed in~\cref{fig:matchplot} the improvement in mismatch for advanced LIGO is only small, but still noticeable and in a quite interesting part of the parameter space. The mismatch will become more pronounced in future detectors since these will be not only more sensitive overall, but also are expected to be especially more sensitive toward lower frequencies. 

Following Ref.~\cite{baird-2012} one can associate the calculated matches with confidence regions of distinguishability of the two waveform templates. These computations are done under the assumptions of Gaussian and stationary noise using the Fisher Matrix formalism which is valid only for the case of high signal-to-noise ratio events~\cite{vallisneri-2007}. We also note that this criteria is necessary but not a sufficient condition as shown in Ref.~\cite{purrer-2019}. Although the results presented here are robust, a full parameter estimation study can vary the confidence intervals where the memory contribution occurs. Since we have three parameters that define the templates, namely total mass, mass ratio and tidal deformability of the NS, a match of 0.97 corresponds to a 90\% confidence region at a signal-to-noise ratio (SNR) of~10. In~\cref{fig:matchplot} a yellow line indicates the contour of 0.97 match, below which the BBH and NSBH templates can be distinguished well if the part of the signal above 500~Hz has SNR of $\ge 10$. From the figure we note that the cosmic explorer performs better than the Einstein Telescope, this can be attributed to the better  sensitivity of Cosmic Explorer in the midfrequency region of $10-100$~Hz.

\section{Interpretation of the detected events} \label{sec:events}

In this section we narrow down the discussion to four detected events which are of interest to us, namely GW190814, GW200105, GW200115 and GW190425. As discussed in Sec.~\ref{sec:tidal}, for all of these binary coalescences at least one of their progenitors has a mass in the range where it can be a NS ($\lesssim 3 \Msun$). However, since no EM counterpart was detected or tidal deformation measured, an ambiguity as to the nature of the compact object remains. 

GW190814 had companions with masses $23.2^{+1.1}_{-1.0}~M_{\odot}$ and $2.59^{+0.08}_{-0.09}~M_{\odot}$ in the 90\% credible intervals. The nature of the lighter companion is therefore particularly ambiguous. As discussed in Sec.~\ref{sec:memory}, the amplitude of the memory signal decays with higher mass ratios. Additionally, the high total mass causes the merger to happen slower (occur at lower frequencies) and likewise the memory build-up which therefore almost has no amplitude for frequencies above 10 Hz. Thus memory cannot contribute in gaining supplementary information about such events.

GW200105 and GW200115 were identified as NSBH mergers based on the mass estimate of the lighter companion in the binary as $\{ 1.9^{+0.3}_{-0.2}~M_{\odot},\, 1.5^{+0.7}_{-0.3}~M_{\odot} \}$ respectively. The masses of the heavier companions were estimated as $\{ 8.9^{+1.2}_{-1.5}~M_{\odot},\, 5.7^{+1.8}_{-2.1}~M_{\odot} \}$ respectively. The spin magnitude of the heavier companion for GW200105 was estimated to be below 0.23 at the 90\% credible level and for GW200115 the spin had support for the negative spin projection with respect to the orbital angular momentum plane at 88\% probability~\cite{LIGO-NSBH}. As expected, given the high mass ratio, the tidal deformation and spin of the secondary object were difficult to measure and could not confirm the presence of a NS. Tidal disruption for these events was deemed to be unlikely, due to the high mass ratio and low positive aligned spin components~\cite{LIGO-NSBH,zhu-2021}. The inclusion of memory for these events is not particularly helpful here. As can be seen from Fig.~\ref{fig:matchplot}, for mass ratios greater than 3 the contribution of memory is negligible for the LIGO detectors at its design sensitivity. However, it should be noted that for the future generation of GW detectors, especially Cosmic Explorer, memory will be a necessary feature for distinguishing such events with high confidence as NSBH or BBH binaries.  

GW190425 was observed by LIGO Livingston detector and was identified as a compact binary coalescence event with a total mass of $3.4^{+0.3}_{-0.1}~\Msun$. This is the event of primary interest for this work: it is considerably more massive than any other BNS system known, of the 19 Galactic BNSs the most massive has a total mass of $2.89~\Msun$~\cite{farrow-2019,agazie-2021}. Nonetheless, the individual components are below the highest precise NS mass measurement of $2.14~\Msun$~\cite{cromartie-2019}. Still an ambiguity in determining the nature of the components remains, since no matter effects were visible in the GW signal nor was an associated electromagnetic observation reported~\cite{coughlin-2019,boersma-2021}.

We performed a follow-up search for nonlinear memory on the event GW190425. We employed the transient search and reconstruction algorithm \textit{coherent WaveBursts} (cWB)~\cite{klimenko-2015} which is sensitive to the memory signal~\cite{ebersold-2020}. We utilize the publicly available data of 4096 seconds duration from the LIGO Livingston detector~\cite{gw190425-data,LIGOScientific:2019lzm} where the glitches were modeled and removed using the BayesWave algorithm~\cite{ glitch-model, cornish-2014}. 
We excise the oscillatory maximum likelihood waveform from the data and conduct a follow-up analysis for the detection of memory. We choose a 40 second time window after the frequency of the oscillatory waveform goes beyond 512~Hz (we refer to this as the on-source time). The rest of the data (of about 4000 seconds in duration) is considered for the estimation of the background (we refer to this as the off-source time). Furthermore, we tune our search to look only for the triggers with central frequency below 200~Hz, this is motivated by the fact that the nonlinear memory signal does not contribute above 150~Hz as shown in Ref.~\cite{ebersold-2020}.

In our analysis we find the loudest trigger in the on-source time to have a false-alarm Rate of about 100 per second with off-source time of 4000 seconds, this corresponds to a p-value of $0.39$ which is statistically insignificant. Here we cannot claim that the event GW190425 is either a NSBH or BBH due to the nondetection of memory, since we do not expect to be able to detect memory with the sensitivity of the detector at the time of this event. We only expect to detect memory with p-value $\leq 0.05$ if the progenitor of GW190425 was a BBH and the merger should be at a distance of around 2~Mpc, so orders of magnitude closer to us than what it actually was. It should be noted that the sensitivity to our algorithm in this case is especially deteriorated as this event was a single detector event and the false-alarms are more difficult to remove.

\section{Conclusion} \label{sec:conclusion}
In this work we have provided yet another nonfungible property of memory; to distinguish tidally disrupted events of NSBH systems from the BBH systems. Even though the memory signal is much weaker compared to the oscillatory signal it contributes in the part of the parameter space where, for the oscillatory signal, distinguishing tidally disrupted NSBH events from BBH systems is most difficult (equal masses and relatively small tidal deformability parameter). Memory also aids the most for edge-on systems, where a potential EM counterpart is less likely to be observed.

We have also quantified the effect of adding memory to waveform models for current and future generation of detectors in terms of (mis-)match and provided evidence that memory contributes in enlarging the parameter space for the distinguishablity of the tidally disrupted NSBH systems from BBH systems. We discuss the consequences of adding memory for the recently detected events in particular GW190425, where we found that for current sensitivity of detectors the use of memory to determine the nature of the lighter object is inconclusive. 

The present work motivates the future development of models with the inclusion of memory, as this provides several advantages not limited to only NSBH systems. For example since memory signal has a very different dependence on the orientation of the binary than the oscillatory signal, it can potentially contribute to better measurement of the inclination angle especially for equal mass systems.

\section{Acknowledgments}
The authors are grateful to Francesco Pannarale for his meticulous reading and comments on the early version of the draft. We thank Maria Haney, Giovanni A. Prodi and David Nichols for insightful discussions and comments. 
S. T. and M. E. are supported by the Swiss National Science Foundation (SNSF) grant No. 200020-182047. E. Z. H is supported by SNSF grant No. IZCOZ0-189876. \\
This research has made use of data, software and/or web tools obtained from the Gravitational Wave Open Science Center (https://www.gw-openscience.org/ ), a service of LIGO Laboratory, the LIGO Scientific Collaboration and the Virgo Collaboration. LIGO Laboratory and Advanced LIGO are funded by the United States National Science Foundation (NSF) as well as the Science and Technology Facilities Council (STFC) of the United Kingdom, the Max-Planck-Society (MPS), and the State of Niedersachsen/Germany for support of the construction of Advanced LIGO and construction and operation of the GEO600 detector. Additional support for Advanced LIGO was provided by the Australian Research Council. Virgo is funded, through the European Gravitational Observatory (EGO), by the French Centre National de Recherche Scientifique (CNRS), the Italian Istituto Nazionale di Fisica Nucleare (INFN) and the Dutch Nikhef, with contributions by institutions from Belgium, Germany, Greece, Hungary, Ireland, Japan, Monaco, Poland, Portugal, Spain.
The authors are grateful for computational resources provided by the LIGO Laboratory and supported by National Science Foundation Grants PHY-0757058 and PHY-0823459. We also acknowledge the provision of computational resources from the Physik-Institut of the University of Zurich. 

\bibliographystyle{apsrev4-1}
\bibliography{references}

\end{document}